\documentclass[preprint,12pt]{elsarticle}
\usepackage{url}
\usepackage{subfig}
\makeatletter
\def\ps@pprintTitle{%
 \let\@oddhead\@empty
 \let\@evenhead\@empty
 \def\@oddfoot{\centerline{\thepage}}%
 \let\@evenfoot\@oddfoot}
\makeatother
\usepackage{amssymb}
\usepackage{hyperref}

\journal{}

\begin{document}

\begin{frontmatter}


%% \Affiliation{organization={},

\title{Automated Atrial Fibrillation Classification Based on Denoising Stacked Autoencoder and Optimized Deep Network}

%% \AFibfiliation[label1]{organization={},
%% \AFibfiliation[label2]{organization={},
\date{}
\author[inst1]{Prateek Singh}   

\affiliation[inst1]{organization={Electrical Engineering Department},%Department and Organization
            addressline={Indian Institute of Technology, Roorkee}, 
            %city={Roorkee},
            postcode={247667}, 
            state={Uttrakhand},
            country={India}}

\author[inst1]{Dr. Ambalika Sharma}
\author[inst2]{Dr. Shreesha Maiya}

\affiliation[inst2]{organization={Leicester Royal Infirmary and Glenfield Hospital},%Department and Organization
            addressline={Infirmary Square }, 
            city={Leicester},
            postcode={LE1 5WW}, 
            state={United Kingdom},
            country={England}}

\begin{abstract}
\paragraph*{Background and Objective}
The incidences of atrial fibrillation (AFib) are increasing at a daunting rate worldwide. For the early detection of the risk of AFib, we have developed an automatic detection system based on deep neural networks. The early detection may help survive many patients by curative measures in clinical settings. Eventually, this may reduce the mortality rate due to AFib. 
\paragraph*{Methods}
For achieving better classification, it is mandatory to have good pre-processing of physiological signals. Keeping this in mind, we have proposed a two-fold study. First, an end-to-end model is proposed to denoise the electrocardiogram signals using denoising autoencoders (DAE). To achieve denoising, we have used three networks including, convolutional neural network (CNN), dense neural network (DNN), and recurrent neural networks (RNN). Compared the three models and CNN based DAE performance is found to be better than the other two. Therefore, the signals denoised by the CNN based DAE were used to train the deep neural networks for classification. Three neural networks' performance has been evaluated using accuracy, specificity, sensitivity, and signal to noise ratio (SNR) as the evaluation criteria. 
\paragraph*{Results}
 The proposed end-to-end deep learning model for  detecting atrial fibrillation in this study has achieved an accuracy rate of 99.20\%, a specificity of 99.50\%, a sensitivity of 99.50\%, and a true positive rate of 99.00\%. The average accuracy of the algorithms we compared is 96.26\%, and our algorithm's accuracy is 3.2\% higher than this average of the other algorithms. The CNN classification network performed better as compared to the other two. Additionally, the model is computationally efficient for real-time applications, and it takes approx 1.3 seconds to process 24 hours ECG signal. The proposed model was also tested on unseen dataset with different proportions of arrhythmias to examine the model's robustness, which resulted in 99.10\% of recall and 98.50\% of precision. 

%\paragraph*{Conclusion}
%The proposed algorithm seems promising in identifying the risk of AFib as evaluated with retrospective data. We suggest an evaluation of the algorithm with a more diverse dataset to develop a robust diagnostic method. This early detection may help clinicians to improve survival rate.  
\end{abstract}


% \begin{highlights}
% \item Proposed a deep learning scheme for automatic detection of atrial fibrillation. 
% \item Three Denoising Autoencoders
% are proposed and compared for denoising ECG signals.
% \item Three end-to-end models are trained and compared for atrial fibrillation classification. 

% \end{highlights}


\end{frontmatter}


\section{INTRODUCTION}
\label{sec:sample1}

Atrial Fibrillation (AFib) is an arrhythmia due to the irregular and rapid beating of the heart's upper chamber (atria). AFib is characterized by highly variable ventricular beat intervals and a rhythm disorder that differs from normal sinus rhythm (NSR) in RR intervals. AFib is the most common type of sustained cardiac arrhythmia worldwide. According to the world health organization (WHO) data, millions of deaths are attributed to Cardiovascular diseases (CVDs) each year \cite{mathers2004global}. CVDs are the leading cause of sudden cardiac death worldwide. Early detection of AFib can help avoid various heart complications such as atrial thrombosis, stroke and heart failure etc. The most effective and feasible approaches to detecting CVD is an electrocardiography test. The test output is an electrocardiogram (ECG) signal that shows the heart's electrical activity and represents the plot of voltage against time. Medical practitioners visually analyze the ECG signal for diagnosing AFib, which is a time-consuming and tedious process. Therefore, the automatic detection of AFib is required. CVD can be characterized by detecting cardiac abnormalities that open archaic avenues for researchers.  
  
 In the past few decades, the researchers proposed several computer-aided methods to detect AFib and other arrhythmias automatically. Several methods are proposed to enable automation. The primary task for arrhythmia detection is to categorize each heartbeat (typically consisting of a P wave, a QRS complex and a T wave) into a set of predefined classes. Some of them include Feature extraction approach \cite{marinho2019novel}, wavelet-based techniques \cite{zhao2005ecg}, support vector machine-based techniques\cite{luz2016ecg}, hidden Markov model-based techniques \cite{coast1990approach} and Machine learning (ML) models, and Deep Learning (DL) approaches \cite{garcia2017inter} \cite{rajpurkar2017cardiologist} \cite{shi2020automated}. ML and DL models show promising results in classifying AFib. Models available in literature often failed to provide adequate sensitivity and specificity because of inter-patient variability, highly skewed class-wise performance and noise Etc. Therefore, an improved approach to segregate different heartbeats based on arrhythmia is the need for time. 
  
  Primary potential noise sources are device power interference, baseline drift, muscle noise, electrode contact noise. Among these motion artifacts, caused by muscle movements, can be mistakenly registered as arrhythmia \cite{berkaya2018survey}. Therefore, proper pre-processing is an essential step before analyzing the ECG signal. Several approaches are proposed to denoise ECG signals, the majority of them are traditional methods that are based on parameters that are highly vulnerable to noise, such as fixed filters like finite impulse response filters (FIR)\cite{van1985removal} and infinite impulse response filters(IIR)\cite{chavan2008suppression}. Wavelet methods\cite{martinez2004wavelet}\cite{li1995detection}\cite{addison2005wavelet}, adaptive filtering \cite{singh2017adaptive}\cite{chandrakar2012denoising}, and empirical mode decomposition (EMD) \cite{blanco2008ecg}\cite{chacko2012denoising} are the other set of methods which requires high computational resources. In the modern era, mobile-based ECG devices and patches are introduced that require less complex and high yield algorithms. One such method for filtering ECG is Denoising Autoencoders (DAE) that has shown better filtering capabilities than other techniques because of their powerful nonlinear mapping capabilities.\cite{chiang2019noise} proposed fully convolution network (FCN)\cite{long2015fully} based DAE for signal denoising. FCN network used is of 13 layers $SNR_imp$ achieved was 15.49db on MIT-BIH database signal, which was corrupted by noise from MIT-BIH-NST database \cite{moody1984noise}.
  
  A variety of methods have been proposed to improve classification accuracy. ML and DL approaches were explored in the recent past. Most popular ML techniques are Support Vector Machines (SVM) and decision tree \cite{park2014pchd}, feature engineering \cite{ledezma2019modeling} and spectral analysis\cite{kotriwar2018higher}, though this often limits the classification scope. Apart from this DL based architectures are popular and shown state-of-the-art results. Different Dense Neural Network (DNN), Convolution Neural Network (CNN) \cite{Fukushima1982NeocognitronAS} and Long Short Term Memory (LSTM) \cite{hochreiter1997long} based architectures are proposed for better classification \cite{isin2017cardiac} \cite{nurmaini2019automated}.DL based classification shown better accuracy for AFib detection\cite{faust2018automated}\cite{mousavi2020han} \cite{andersen2019deep}.  
 
 To the best of the author's knowledge, no study is conducted to analyse the ECG denoising effect on deep networks used for classification. Therefore, the authors have proposed improved deep learning-based denoising autoencoders using DNN, CNN, and RNN to denoise ECG signals and compared their performances. Then, the filtered signals are given to a deep learning model for AFib classification to study the effect of different denoising autoencoders on classification performance, which have been shown to improve performance without the significant increase in model complexity typically seen in DL architectures.
 
 In this study, a comparison between the proposed three denoising autoencoders (DAE) architecture is made, and their performance was studied on different noises. Then, the filtered signals obtained from DAEs are used to classify Atrial fibrillation (Afib) and studied the effect of robust filtering on classification using the proposed CNN network. Deep architectures are optimized to provide efficient and robust performance, thereby optimizing resource utilization. The methodology is shown in Figure 1.

\begin{figure}[!h]\label{fig1}
\centering{\includegraphics[width=12cm,scale=0.2]{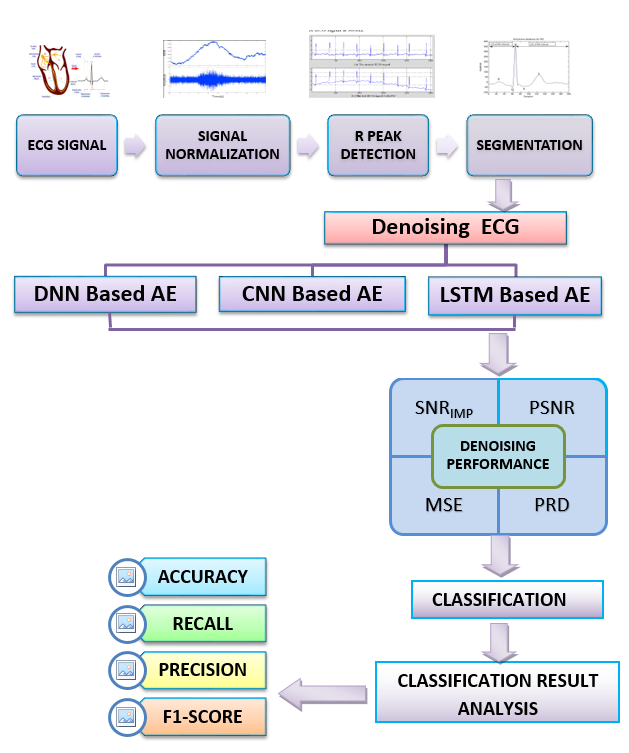}}
\caption{Flowchart of Proposed Methodology}
\end{figure}

\section{MATERIALS AND METHODS:}
In this section, the database and pre-processing methods are explained, for detection of AFib. All pre-processing steps and model description are discussed, along with the computational complexity.
\subsection{DATASET Used :}
Neural Network was trained on the MIT-BIH Atrial fibrillation database (AFDB) and  MIT-BIH Normal Sinus Rhythm Database (NSRDB) \cite{goldberger2000physiobank} for AFib classification. AFDB includes twenty-five 10 hours of ECG recordings sampled at 250Hz of which raw ECG of two records (“00735” and “03665”) are not available, and two records (“04936” and “05091”) include many incorrect reference annotations. So, only 21 records are taken for this study. These recordings come with manually prepared rhythm annotation files. In the database, if one or more beats are showing sign of AFib then the beat sequence are classified as AFib(mostly paroxysmal), and all other beats are classifies as normal. The database has 1221574 beats, out of which 519687 are Afib. NSRDB includes 18 long-term ECG recordings.No arrhythmia beats are available, except sporadic ectopy. \\\
 The MIT-BIH Noise Stress Test Database (NSTDB)\cite{moody1984noise} is used to test the denoising performance of DAEs . The database includes noise records (`bw', `em', and `ma') that can be added to ECG records to create noise stress test records.  It also contains 12 sample noise stress test records generated by `nst', by adding `em' noise to MIT-BIH Arrhythmia Database records 118 and 119. For convenience, reference annotation files are included here; all are copies of the original reference annotation files for records 118 and 119. Here 'bw' represents baseline wander, 'em' represents electrode motion and 'ma' represents muscle artifacts. 

\begin{table}[h]
\caption{Database Used} 
\centering 
\begin{tabular}{l c c rrrrrrr} 
\hline\hline 
 Databases &Time,min & AF,records &Normal,records & Sampling,Hz 
\\ [0.05ex]
\hline 
MIT-BIH AFDB &600 &21 &0 & $250$  \\[0.5ex]
MIT-BIH NSRDB& 1500&0 & 18 & $128$  \\[0.5ex]
Total& 2100&21 & 18 & $-$  \\[0.5ex]
\hline % inserts single-line
\end{tabular}
\label{tab:PPer}
\end{table}

\subsection{Pre-processing}

 The experiment was conducted on the Keras with Google Tensorflow 2.3 backend Deep Learning Library on python 3.7 software. The computer used has Intel Core i7-8700 3.20GHz CPU, 16GB memory and 4GB NVIDIA GeForce GTX 1070 graphics card with Cuda libraries. The input to the proposed denoising autoencoders is ECG signal and noise. The ECG signal is pre-processed before giving it to the proposed model as explained:
 \begin{enumerate}
     \item The ECG signals from the NSRDB is resampled to 250Hz to make them compatible to AFDB database.
     \item Then the signal is normalized between 1 and -1  
     \item R peaks are detected using Pan Tompkins algorithm\cite{pan1985real}.
     \item The continuous ECG signals are divided into a sequence of heartbeats and splitted into normal and AFib based on the annotation files. Window of 1.2 sec is taken as the normal range for RR interval is 0.6-1.2 seconds \url{https://emedicine.medscape.com/article/2172196-overview} as AFib beats RR intervals are always smaller than the normal beats.The segmented windows are zero padded as Afib has less number of samples as shown in Figure 2.
     \item Data prepared is shuffled and splitted into test, train and validation data in the ratio of 15:75:10 percentage respectively.
     \item To remove the affect of imbalanced data 30,000 samples for AFib and non-AFib is chosen randomly, totally 60,000 samples.
     \end{enumerate}
  
\begin{figure}[!h]\label{fig1}
\centering{\includegraphics[width=15cm,scale=0.2]{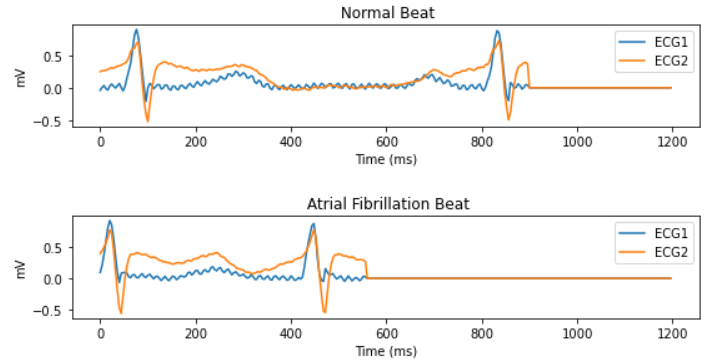}}
    \caption{Normal and AFib sample window}
\end{figure}
 The performance of three DAE architectures was analysed by introducing Additive white Gaussian noise (AWGN), which is an addition of random noises that occur in nature with different decibel(db) scale. The model was also trained on MIT-BIH AFDB database signals after introducing the real noises from the MIT-BIH NST database to adapt to most real noises. These datasets signals are also re-sampled to 250 Hz to match our model specifications.

 \section{Denoising Autoencoders}
DAE architecture is divided into three layers, namely input, encoding and decoding layer. In DAE, input and output should have the same dimension as we are constructing the input. Denoising Autoencoders(DAEs) with Dense Neural Network (DNN), Convolution Neural Network (CNN) \cite{Fukushima1982NeocognitronAS} and Long Short Term Memory (LSTM) \cite{hochreiter1997long} were used for testing of noise filtering capabilities in ECG. DAE model are shown in figure 3. 
\begin{figure}[!h]\label{fig1}
\centering{\includegraphics[width=15cm]{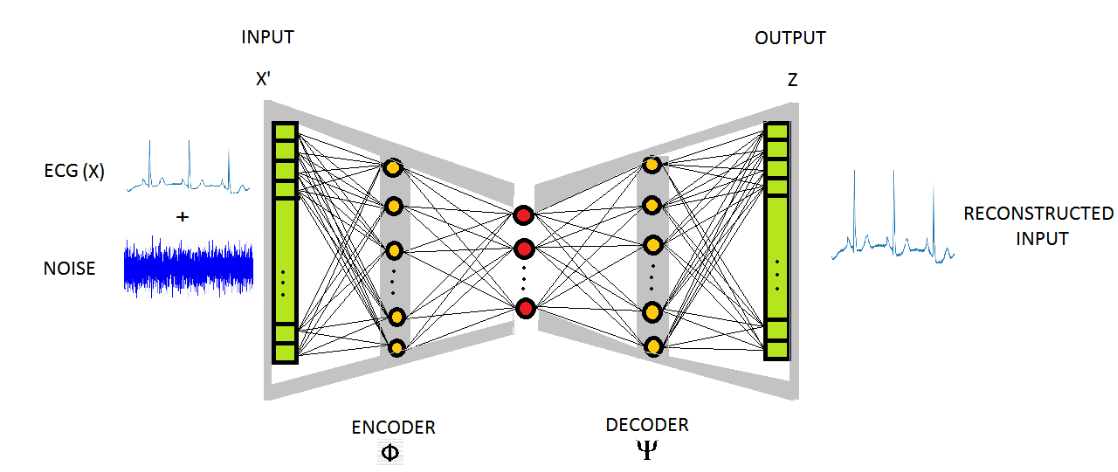}}
\caption{Denoising Autoencoder}
\end{figure}

\begin{equation} 
 \theta  \psi = argmin (\frac{1}{n}  \sum_{i=1}^{n} L(x_i,z_i))
 \end{equation}\\
 
 \begin{equation}
     =  argmin (\frac{1}{n}  \sum_{i=1}^{n} L(x_i,g_ \psi (f_\theta(x_i)))
 \end{equation}

where L is a cost function, the mean squared
error is used as $(MSE): L({x_i}, z_i) =[\![x_i - z_i]\!]^2\\$

AEs are usually used for clean signals and to denoise the noisy signals, denoising AEs (DAE) are introduced by Vincent et al.\cite{vincent2008extracting} where noise is added to the input data and designed such that more robust features can be extracted than the AE algorithm. Figure 1 shows a schematic of DAE. Input vector $x \in {\Bbb [0,1]}^d$ is given to the system, where d is the input vector's dimension. The initial input x is corrupted to $\tilde{x}$ by a stochastic mapping $\tilde{x}  \sim {C(\tilde{x}|x)}$, which partially destroys the input data, as per destruction rate. The algorithm uses the corrupted $\tilde{x}$ as input data and then maps it to the corresponding $y_i$ and ultimately to its reconstruction $z_i$.\\
The encoder converts $\tilde{x}$ to y, that represents partial information. This is a non-linear transformation as $y = f_\theta (x) = \sigma (W \tilde{x} + b)$. 
The decoder converts y back to reconstructed data. This is a non-linear transform as
%  $\hat{x} = g_{\psi} {(y)} = \sigma ({W^'}y + {b^'})$.  $\sigma$ represents the input   $\tilde{x}$ in the feature space. 
 
  Minimizing the objective function as shown in equation 2 is the main objective that can be achieved by training parameters of DAE. 
\begin{equation} 
  \theta  \psi = argmin (\frac{1}{n}  \sum_{i=1}^{n} L(x_i,g_ \psi (f_\theta(\tilde{x_i}))) 
 \end{equation}

In this paper, noisy ECG signals are used to train DAEs before the automated feature extraction for classification, and their denoising performance are tested statistically using signal to noise ratio $(SNR_i)$, mean square error (MSE), Peak to signal noise ratio$(PSNR_i)$,percent root mean square difference (PRD).\\

\subsection{CNN Based DAE Architecture}
\begin{figure}[!h]\label{fig1}
 \centering{\includegraphics[width=15cm,scale=0.2]{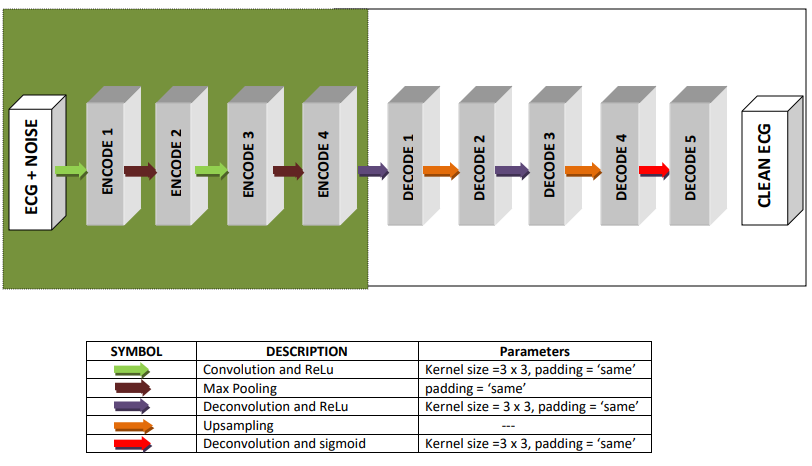}}
     \caption{Architecture of CNN based DAE}
 \end{figure}
The proposed CNN based DAE network is represented in Figure 4. The encoder of convolutional layers and decoder of symmetrical deconvolutional layers together form the CNN network. ECG signal added with noise is received as an input by the network, and the network output is a clean ECG signal. The convolutional layer is a feature extraction layer that extracts ECG features while performing denoising of ECG. Subsequently, the deconvolutional layer recovers the signal's original details by decoding the ECG abstraction performed by the convolutional layer. As ECG signals are very noisy; therefore, a large filter patch is required for efficient results. Large patch size will help to extract more context information from the larger region of the signal. As the patch size increases, the receptor field of the network will also increase. To increase the receptive field, it is often recommended to use the deeper network, but that will increase the computational cost. Therefore, we chose two convolutional and two deconvolutional layers for our model to give optimal results without being too deep. Since our data is temporal, we adopt the 1D convolution and deconvolution layers. Pooling operation is not generally performed in denoising tasks as we need full signal details. But in our case, the pooling operation will help increase the receptive field, so we used max-pooling with the stride of two. The max-pooling layer is used for reducing the convoluted feature's spatial size, which will lead to lessening the computational cost and control overfitting. The 3*3 kernel size is chosen as it was large enough to include sufficient signal information. The "same" padding for optimizing the output size. For non-linearity after each layer, the ReLu activation function with a slope of 0.3 is utilized. Early stopping is also utilized in model training. A batch size of 32 is taken for training.

\subsection{DNN Based DAE Architecture}
\begin{figure}[!h]\label{fig1}
 \centering{\includegraphics[width=14cm,scale=0.2]{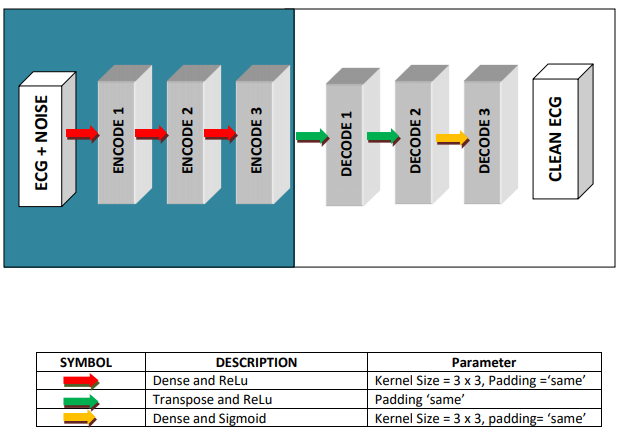}}
     \caption{Architecture of DNN based DAE}
 \end{figure}
 The DNN performs the encoding and decoding of DNN based DAE. A noisy ECG signal is received as input by the network, and the network output is a denoised ECG signal. Three dense layers are used for encoding, and three dense layers are used for decoding. A ReLu activation layer follows each dense layers. The last dense layer is followed by sigmoid activation, as shown in figure 5. 

\subsection {RNN Based DAE Architecture}
RNN architecture is preferred for sequential data. The most widely used RNN architecture is LSTMs. LSTM consists of memory blocks and memory cells, along with gate units \cite{gers1999learning}. They are designed in such a way that they can remember values at arbitrary intervals. The proposed architecture of the LSTM based DAE model is described in figure 6. The LSTM performs the encoding and decoding of LSTM based DAE. Two LSTM layers are used for encoding, and a repeat vector and an LSTM layer is used in decoding, as shown in figure 6.  Here, the LSTM encoder extracts the ECG signals' features while the decoder converts the feature maps to the output. Parameters of the encoders and decoders are computed by using an unsupervised training process. 

\begin{figure}[!h]\label{fig1}
 \centering{\includegraphics[width=11cm,scale=0.2]{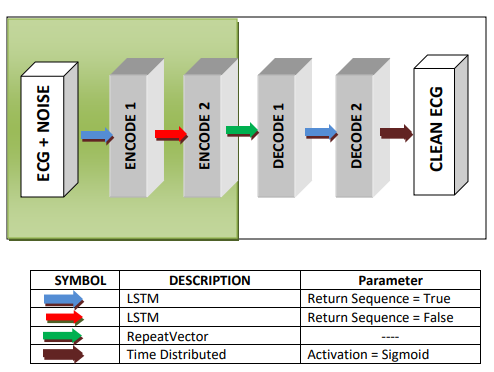}}
     \caption{Architecture of RNN(LSTM) based DAE}
 \end{figure}

\subsection{Network Training}
For training, the network means squared error loss was minimized, which is defined in equation 3. For all three DAE network, Adam algorithm \cite{kingma2015adam} was selected for optimizing, and the learning rate was $1 e^{-5}$ for DNN based DAE and CNN based DAE and $1 e^{-3}$ for LSTM based DAE. The comparison of results is shown in Table 5. CNN based DAE outperformed the other two. So we took the output of CNN based DAE for the classification model.

\section{Classification}
For classification, we have trained and compared three proposed architectures based on CNN, DNN and RNN. The clean signal from the CNN based DAE model is given input to the classification models.    
\subsection{DNN Model for Classification}
 The DNN model contains seven dense layers with a ReLu activation function and a softmax output layer.  Model is trained to minimize categorical cross-entropy loss \cite{zhang2018generalized}. Adam algorithm is used to optimize the model. A Drop-out layer of 0.3, i.e. 30 \%, is used before three dense layers to avoid overfitting. A batch size of 64 is taken, and the model got trained for 50 epochs. The confusion matrix of classification is shown in figure 11, and the classification report is shown in Table 2.
 
 \begin{table}[h]
\caption{DNN architecture} % title name of the table
\centering % centering table
\begin{tabular}{l c c rrrrrrr} % creating 10 columns
\hline\hline % inserting double-line
 Layer &Type&Output Shape & Number of Parameters 
\\ [0.05ex]
\hline % inserts single-line
% Entering 1st row
1 & Dense(ReLu)&  (None,1024) & 308224   \\[1.0ex]
2 & Dropout&(None,1024) & 0   \\[1.0ex]
3 & Dense(ReLu)&  (None,1024) & 1049600 \\[1.0ex]
4 & Dropout&(None,1024) & 0  \\[1.0ex]
5 & dense(ReLu)&(None,512) & 524800  \\[1.0ex]
6 & Dense(ReLu)&(None,128) & 65664 \\[1.0ex]
7 & Dense(ReLu)&(None,64) & 8256 \\[1.0ex]
7 & Dense(Sigmoid)&(None,2) & 130 \\[1.0ex]
%\raisebox{0.1ex}{Total} & \raisebox{0.1ex}{2100} \\[0.1ex]
% [1ex] adds vertical space
\hline % inserts single-line
\end{tabular}
\label{tab:PPer}
\end{table}
 \subsection{CNN Model for Classification}
The CNN model contains three convolution layers, three max-pooling layers, three batch normalization layers, one dropout layer, one fully connected layer and three dense layers shown in Table 3. A batch normalization layer follows each Convolution layer. Batch normalization standardizes the inputs to a layer for each mini-batch. This has the impact of stabilizing the learning process and drastically decreasing the number of training epochs required to train deep neural networks. The batch Normalization layer is followed by the Max pooling layer, which helps control the overfitting and decrease the computational cost. Each max-pooling layer has a pool size of two, a stride of two, and the 'same' padding. To tackle non-linearity ReLu activation function is utilized after every convolution layer and dense layer. For training, networks categorical cross-entropy loss is minimized. Adam algorithm was used to optimize the network. A batch size of 128 is chosen, and the network got trained for 16 epochs.  
\begin{table}[h]
\caption{CNN architecture} % title name of the table
\centering % centering table
\begin{tabular}{l c c rrrrrrr} % creating 10 columns
\hline\hline % inserting double-line
 Layer &Type&Output Shape & Number of Parameters 
\\ [0.05ex]
\hline % inserts single-line
% Entering 1st row
    1 & Conv1D(ReLu) &  (None, 296, 64) & 384   \\[1.0ex]
2 & Batch Normalization&(None, 296, 64) & 256   \\[1.0ex]
3 & Max Pooling&(None, 148, 64) & 0 \\[1.0ex]
4 & Conv1D(ReLu) & (None, 146, 64)& 12352  \\[1.0ex]
5 & Batch Normalization&(None, 146, 64) & 256  \\[1.0ex]
6 & Max Pooling&(None, 73, 64) & 0 \\[1.0ex]
7 & Conv1D(ReLu) & (None, 71, 64)& 12352  \\[1.0ex]
8 & Batch Normalization&(None, 71, 64) & 256  \\[1.0ex]
9 & Max Pooling&(None, 36, 64) & 0 \\[1.0ex]
10 & Flatten&(None, 2304) & 0 \\[1.0ex]
11 & Dropout&(None, 2304) & 0 \\[1.0ex]
12 & Dense(ReLu) & (None, 128) & 295040 \\[1.0ex]
13 & Dense(ReLu) & (None, 32) & 4128 \\[1.0ex]
14 & Dense(Sigmoid) & (None, 2) & 66 \\[1.0ex]

%\raisebox{0.1ex}{Total} & \raisebox{0.1ex}{2100} \\[0.1ex]
% [1ex] adds vertical space
\hline % inserts single-line
\end{tabular}
\label{tab:PPer}
\end{table}
\subsection{RNN Model for Classification}
The details of the Bidirectional LSTM network are shown in Table 4. A CNN-BiLSTM approach for the classification of AFib is published recently \cite{wang2020atrial}. Bidirectional LSTM cells are followed by a dropout layer of 0.2. The dense layer is used with the ReLu activation function. The last dense layer gives us the output, and as this is a binary classification, so sigmoid activation is used for the last layer. Thirty-eight thousand three hundred eighty-eight parameters are trained during the process, and 50 epochs are used for training.
\begin{table}[h]
\caption{RNN(Bidirectional LSTM) architecture} % title name of the table
\centering % centering table
\begin{tabular}{l c c rrrrrrr} % creating 10 columns
\hline\hline % inserting double-line
 Layer &Type&Output Shape & Number of Parameters 
\\ [0.05ex]
\hline % inserts single-line
% Entering 1st row
    1 & Input&  (None,300) & -   \\[1.0ex]
2 & Bidirectional(LSTM)&(None,300,64) & 8960   \\[1.0ex]
3 & Bidirectional(LSTM)&(None,64) & 25088 \\[1.0ex]
4 & Dropout&(None,64) & 0  \\[1.0ex]
5 & $dense_1$(ReLu)&(None,64) & 4160  \\[1.0ex]
6 & $dense_2$(Sigmoid)&(None,64) & 130  \\[1.0ex]
%\raisebox{0.1ex}{Total} & \raisebox{0.1ex}{2100} \\[0.1ex]
% [1ex] adds vertical space
\hline % inserts single-line
\end{tabular}
\label{tab:PPer}
\end{table}

\section{Experimental Results }

\subsection{Experimental Results of Proposed Denoising Autoencoders}

 The denoised signals and reconstruction quality of all three proposed DAE, i.e. DNN based DAE, CNN based DAE, and LSTM based DAE, is shown in Figure 7. In Figure 7, a random window is taken from the test set, and AWGN noise of -10 dB is added to the signal, and the signal is passed to the three trained models, i.e. DNN based DAE, CNN based DAE, and RNN based DAE, respectively and the denoised signal is plotted as shown. 
 
 \begin{figure}
\centering
\subfloat %Original Signal
{
  \includegraphics[width=11.5cm]{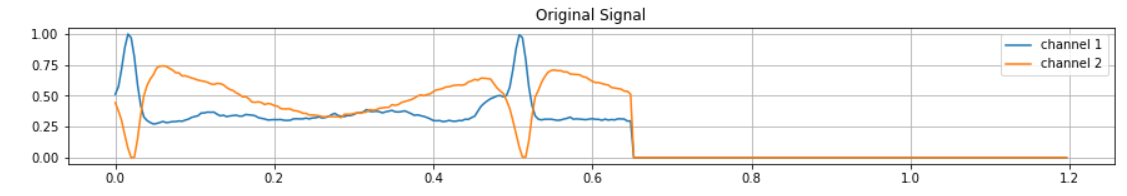}
}
\subfloat %Noisy Signal with -10dB AWGN
{
  \includegraphics[width=11.5cm]{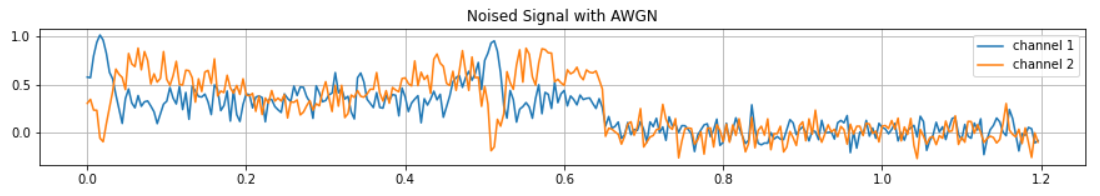}
}
%\hspace{5mm}
\subfloat %DNN Based DAE Output
{
  \includegraphics[width=11.5cm]{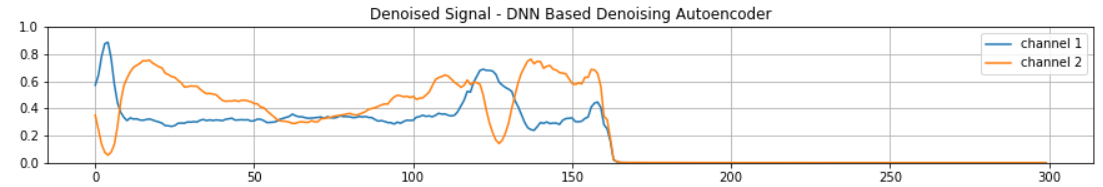}
}
\subfloat %CNN Based DAE Output
{
  \includegraphics[width=11.5cm]{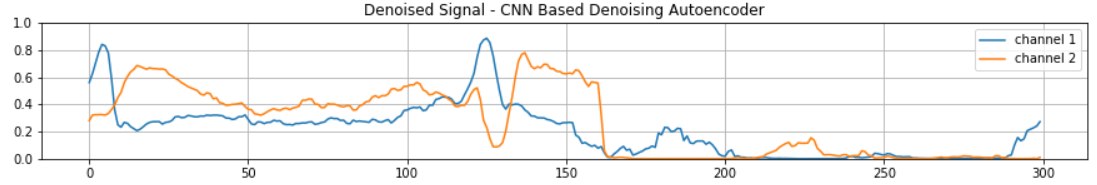}
}
\hspace{0mm}
\subfloat %LSTM Based DAE Output
{   % ???
  \includegraphics[width=11.5cm]{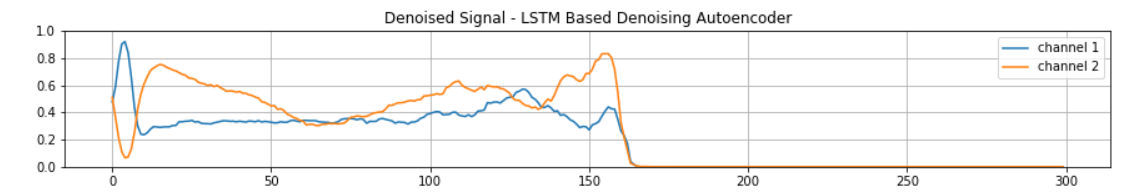}
}

\caption{Comparison of DAE output}
\end{figure}
 Figure 8 represents the violin plot of all three models, which represents the SNR improvement. Further, statistical analysis using signal to noise ratio (SNR), peak signal to noise ratio (PSNR), mean square error (MSE), and percent root mean square difference(PRD) metrics were used to evaluate the denoising results obtained for the test data. Mathematically, 
 
\begin{equation}\label{SNReq}
SNR = \frac{\mu_{signal}}{\sigma_{denoised}}
\end{equation}

\begin{equation}\label{MSEeq}
MSE = \frac{1}{N} \sum{(X - X_{denoised})^2} 
\end{equation}

\begin{equation}\label{MSEeq}
PSNR = 10 \log_{10}\frac{{X_{max}}^2}{MSE}
\end{equation}

\begin{equation}\label{PRDeq}
PRD = \sqrt{\frac{\sum{(X - X_{denoised})^2}}{X^2}} * 100
\end{equation}

where  $\mu_{signal}$ = mean of the signal and
 $\sigma_{denoised}$ = standard deviation of noise
$N$ is the total number of terms for which the error is to be calculated,
$X$ is the original signal.

 \begin{figure}[!h]\label{fig1}
\centering{\includegraphics[width=11cm,scale=0.5]{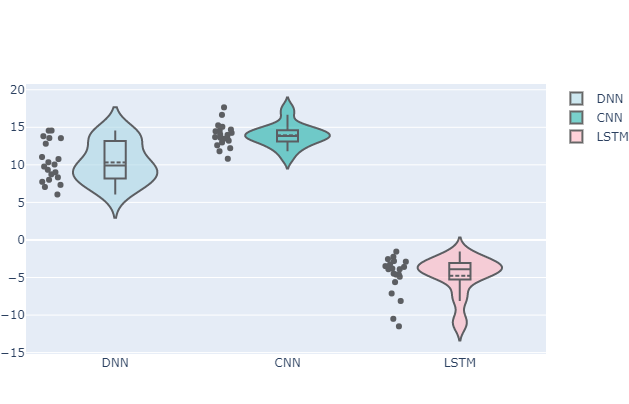}}
     \caption{Violin Plot for SNR improvement at -10db noise}
 \end{figure}

 Table 5 shows the SNR improvement, PSNR, PRD comparison.  It is visible that CNN based DAE outperformed the other two, whereas LSTM based DAE was the worst. As per the performance, we have chosen CNN based DAE output signal for the classification of AFib.

 \begin{table}[h]
\caption{Comparision of DAE statistics} % title name of the table
\centering % centering table
\begin{tabular}{l c c rrrrrrr} % creating 10 columns
\hline\hline % inserting double-line
 DAE &SNR improvement & PSNR &PRD(\%)
\\ [0.05ex]
\hline % inserts single-line
% Entering 1st row
DNN Based DAE &9.82$\pm {2.69}$dB &0.98 &5.8  \\[0.5ex]
%\raisebox{1ex}{MIT-BIH AFDB} & \raisebox{0.1ex}{600}  \\[1ex]
% Entering 2nd row
CNN Based DAE& 13.43 $\pm {1.51}$dB&0.95 & 3.6  \\[0.5ex]
%\raisebox{1ex}{MIT-BIH NSRDB} & \raisebox{0.1ex}{1500} \\[1ex]
% Entering 3rd row
RNN(LSTM) Based DAE& -5.11 $\pm {2.66}$dB&-6.17 & 18.3   \\[0.5ex]
%\raisebox{0.1ex}{Total} & \raisebox{0.1ex}{2100} \\[0.1ex]
% [1ex] adds vertical space
\hline % inserts single-line
\end{tabular}
\label{tab:PPer}
\end{table}

 We have also tested our model's performance by infusing the real noises from the NST database, as shown in Figure 8. We have added all the three noises in equal proportion i.e $33.3\%$ of muscles artifacts 'ma', baseline wander 'bw' and electrode motion 'em'. Passed the noisy signal to our CNN based DAE model and received clean signal as shown in Figure 9, where blue colour represents a noisy signal, red colour represents the original signal and signal in green represents denoised reconstructed ECG signal. 
\begin{figure}[!h]\label{fig1}
 \centering{\includegraphics[width=13cm,scale=0.2]{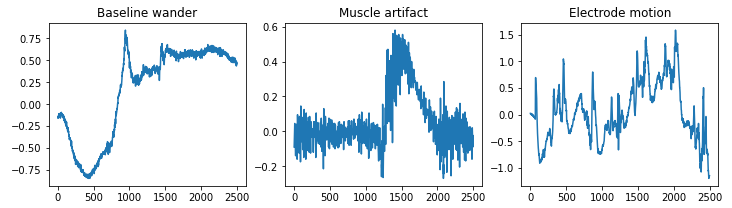}}
    \caption{'bw','ma','em' noise from NST database}
    \end{figure}

\begin{figure}[!h]\label{fig1}
 \centering{\includegraphics[width=14cm,scale=0.2]{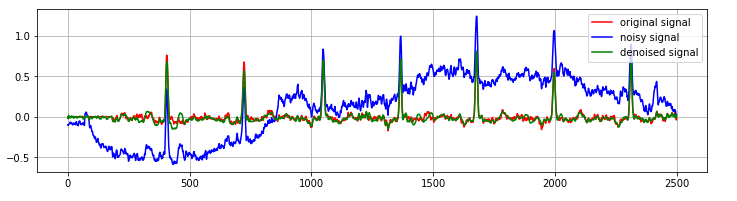}}
     \caption{Denoised ECG using CNN based DAE}
 \end{figure}

\subsection{Experimental Results of Proposed Classification Models}

The classification results of all three models are shown with the confusion matrix's help, as in Figure 11. We have tested a total of 9000 beats which are a mix of AFib and normal beats.  To evaluate the performance of the model, standard statistical measures are performed. Mathematically given as,  

\begin{equation}\label{Accueqn}
Accuracy = \frac{TP+TN}{TP+FP+FN+TN}
\end{equation}
\begin{equation}\label{Accueqn}
Precision = \frac{TP}{TP+FP}
\end{equation}
\begin{equation}\label{Preceqn}
Recall = \frac{TP}{TP+FN}
\end{equation}
\begin{equation}\label{ F1eqn}
F1-Score = 2*\frac{(Recall * Precision) }{(Recall+Precision)}
\end{equation}

where, TP is True Positive, TN is True Negative, FP is False Positive, FN is False Negative.

\begin{figure}[ht]
%\begin{subfigure}%{.5\textwidth}
  \centering
  % include first image
  \includegraphics[width=.4\linewidth]{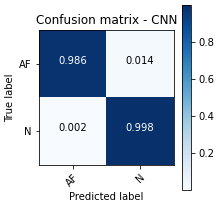}  
  %\caption{Put your sub-caption here}
  \label{fig:sub-first}
%\end{subfigure}
%\begin{subfigure}%{.5\textwidth}
  \centering
  % include second image
  \includegraphics[width=.4\linewidth]{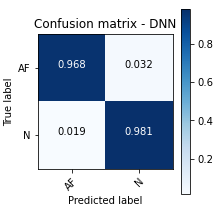}  
  %\caption{Put your sub-caption here}
  \label{fig:sub-second}
%\end{subfigure}
%\begin{subfigure}%{.5\textwidth}
  \centering
  % include second image
  \includegraphics[width=.4\linewidth]{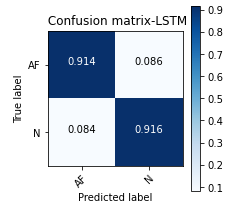}  
  %\caption{Put your sub-caption here}
  \label{fig:sub-second}
%\end{subfigure}
\caption{Confusion matrices for all the proposed methods on the MIT-BIH AFDB database}
\label{fig:fig}
\end{figure}
Hyperparameter tuning of the classification model is done with the help of Google's Keras-tuner library \cite{manaswi2018deep} for optimization. This library helps to pick the optimal set of hyperparameters in the hidden layer and dropout layers. We have provided different values for different parameters: regularization rate of $L2$ regularization technique, number of neurons, dropout rate, and learning rate of Adam optimizer. We chose number of neurons in the range of [$2:20$], dropout rate in the range of [$0.1:0.5$], learning rate of Adam optimizer [$1e^{-2},1e^{-3},1e^{-4},1e^{-5}$].Then, each technique's performance is evaluated by measuring accuracy, precision, recall, and F-Measure. The set of parameters explained in the architectures are tuned from the Keras-tuner library.

CNN classification model performed best for our retrospective data, where we achieved 99.20\% of accuracy ith 99.50\% of recall and 99.50\% of precision. As per authors knowledge, this is the best classification results achieved on AFDB dataset. DNN classification model came second with 97.45 \% of accuracy whereas RNN classification model came last with accuracy of 91.5\% of accuracy.

\begin{table}[h]
\caption{Comparison of the performances of AF detection algorithms that have been validated on MIT-BIH AFDB database.} % title name of the table
\centering % centering table
\begin{tabular}{l c c rrrrrrr} % creating 10 columns
\hline\hline % inserting double-line
 Algorithm &Precision & Recall &F-1 Score & Accuracy
\\ [0.05ex]
\hline % inserts single-line
% Entering 1st row
Asgari, et al. (2015)\cite{asgari2015automatic} & 97.10&97.00 & - & -  \\[1.0ex]
Xia, et al. (2018)\cite{xia2018detecting} & 97.87&98.79 & - & 98.63  \\[1.0ex]
Wang, et al. (2019)\cite{wang2020atrial} & 97.1&97.9 & - & 97.4  \\[1.0ex]
Andersen et al. (2019)\cite{andersen2019deep} & 97.80&98.98 & - & 97.80  \\[1.0ex]
Mousavi et al. (2020)\cite{mousavi2020han} & 98.50&99.80 & - & 98.80  \\[1.0ex]
Petmezas et al. (2021)\cite{petmezas2021automated} & 99.29&97.87 & - & -  \\[1.0ex]
\textbf{Proposed DNN Model}  & 97.50&97.50 & 97.50 & 97.45  \\[1.0ex]
\textbf{Proposed CNN Model}  & 99.50&99.50 & 99.0 & 99.20  \\[1.0ex]
\textbf{Proposed RNN Model}  & 91.50&91.50 & 91.50 & 91.5 \\[1.0ex]
%\raisebox{0.1ex}{Total} & \raisebox{0.1ex}{2100} \\[0.1ex]
% [1ex] adds vertical space
\hline % inserts single-line
\end{tabular}
\label{tab:PPer}
\end{table}

We now wanted to test the model on real patient signals those are unseen to the model and have AFib episodes as well as the normal rhythms. We have created a 5 such signals of varying length and with different distribution of AFib/Normal ratio. The signals are taken from the Long term AF Database\cite{petrutiu2007abrupt} which is sampled at 128Hz and has 84 records, we have chosen random five signals with different distribution of AFib and Normal beats using annotation. The distribution of data is shown in Table 7. 

\begin{table}[h]
\caption{Validation Signal} % title name of the table
\centering % centering table
\begin{tabular}{l c c rrrrrrr} % creating 10 columns
\hline\hline % inserting double-line
Signal &Normal beats & AFib beats & Normal beat (\%) & AFib beat (\%)
\\ [0.05ex]
\hline % inserts single-line
% Entering 1st row
Signal 1 & 256 & 198 & 56.4 & 43.6  \\[0.5ex]
Signal 2 & 471 & 83 & 85 & 15  \\[0.5ex]
Signal 3 & 869 & 192 & 81.9 & 18.1 \\[0.5ex]
Signal 4 & 234 & 15 & 94 & 6  \\[0.5ex]
Signal 5 & 23 & 2 & 92 & 8  \\[0.5ex]
\hline % inserts single-line
\end{tabular}
\label{tab:PPer}
\end{table}

We got 98.8\% of average accuracy for all the five signals with recall of 99.1\% and precision of 98.5\%.

\section{Discussion}

The proposed model achieved state-of-the-art results, indicating that it outperforms the traditional feature engineering techniques. The proposed CNN based DAE model has shown good denoising capabilities, as indicated in Table 5. Previous DL models \cite{andersen2019deep} struggles with a high number of False positives (FPs) due to the noisy signals and lack of proper filtering. Therefore, we focused on proper denoising technique before classification, and tested our DAE model on AWGN noise and most encountered real ECG noises from the MIT-BIH NST database, as shown in Figure 7, 9 and 10. It is visible that our best performing CNN-based DAE achieved an average SNR improvement of 13.43 dB and negligible change in the morphology of ECG signal shown in Figures 7 and 10. 

We then compared three proposed classification models based on DNN, CNN and RNN; architectures are shown in Table 2, 3 and 4, respectively. Table 6 shows the performance of all three classification models on the AFDB database. Here, the CNN classification model outperforms the other two with an accuracy of 99.20\%, and the model achieved 99.50\% of precision and recall with 99\% of F1-score. We further validated our CNN classification model to investigate false positives (FPs). We have chosen the NSRDB, which contains 18 recordings and no AFib episodes are present in the dataset. It will give us a good intuition on FPs. AFDB database shows false positives of 0.83\%, and when it was tested on the NSRDB database, it was 1.12\%, i.e. we found an increase of 0.29\% in false positives, which suggests further investigation. Inspecting raw signals suggest that our R peak detection algorithm failed to detect some of the R peaks where the signals suffer from multiple noises. Therefore, we suggest using signal quality indices to remove high noise segments from the analysis and hence not increase in false-positive rates.  

Again, when we have tested our end-to-end model on unseen data with different proportion of AFib beats, we got 98.8\% of average accuracy for all the five signals with recall of 99.1\% and precision of 98.5\%. So, there was an increase in FPs due to noisy segments which can be resolved by defining proper signal quality indices (SQIs). The proposed algorithm seems promising in identifying the risk of AFib and can be used for real time applications after validating it on more diverse dataset and defining some SQIs. 

\section{Conclusions}
The performance of DAE was compared with performance metrics such as SNR, PSNR and PRD. Among the three networks used, CNN-based DAE performs better with  SNR, PSNR and PRD values of 13.43, 0.95, and 3.6\%, respectively. This indicates that denoising capability and reconstruction quality of the CNN network is better among all. The classification performed with CNN shows the highest accuracy of 99.20 \% with 99.50 \% recall, 99.50 \% precision and 99\% of F1-score. To the best of the authors' knowledge, this is the highest achieved performance reported in the literature. The proposed algorithm seems promising in identifying the risk of AFib as evaluated with retrospective data. We suggest an evaluation of the algorithm with a more diverse dataset to develop a robust diagnostic method. This early detection may help clinicians to improve survival rate.

%appendix~\ref{sec:sample:appendix}.

%\appendix

%\section{Sample Appendix Section}

 \bibliographystyle{elsarticle-num} 
 \bibliography{cas-refs}





% \end{thebibliography}
\end{document}